# Women's Participation in Computing: Evolving Research Methods

Thomas J. Misa

Submitted February 2024; revised July 2024

My 2022 talk for the ACM History Committee's SIG Heritage Seminar on "Why SIG History Matters: New Data on Gender Bias in ACM's Founding SIGs 1970–2000" presented data describing women's participation as research-article authors in 13 early ACM SIGs, finding significant growth in women's participation across the timeframe 1970–2000 and, additionally, remarkable differences in women's participation between the SIGs.[1] That presentation built on several earlier publications where I have developed a research method for assessing the number of women computer scientists that [a] are chronologically prior to the availability of the Bureau of Labor Statistics (BLS) data on women in the IT workforce; and [b] permit focused investigation of varied sub-fields within computing. This present report seeks to connect my earlier articles, and my evolving research method, to that 2022 ACM SIG Heritage presentation. It also outlines some of the choices and considerations I've made in developing and refining "mixed methods" research (using both quantitative and qualitative approaches, described below) as well as extensions of the research I am currently exploring.

---

[1] See ACM History Committee's "Capturing Hidden ACM History" (30 September–1 October 2022) at history.acm.org/seminar-capturing-hidden-acm-history/ and archived at web.archive.org/web/20220928184755/https://history.acm.org/seminar-capturing-hidden-acm-history/

Women's Participation in Computing: Evolving Research Methods (version July 2024)

Figure 1: Men and women in early ACM membership (May 1948)

Note Jean Bartik and Gertrude Blanch, well known women in the profession; gender label "Jr."; but e.g. "Prof. R.C. Archibald" is not gender-identified. *Source:* Margaret R. Fox papers, box 2, folder 9, Charles Babbage Institute 45 at purl.umn.edu/41420.

I started this line of research when Jeff Yost, my colleague at the Charles Babbage Institute (now its director), remarked to me that the records of the IBM user group SHARE (founded in 1955) had extensive lists of the attendees of its twice-yearly conferences – and that these lists contained many <u>first</u> names. On inspection, I found there were a lot of Charles, Harolds, Roberts and other men's names but also a significant number of Marys, Elizabeths, and other women's names. CBI collections also document early ACM members (see figure 1) as well as other significant computing conferences in the 1950s. CBI holds archival records not only of SHARE but also of Univac USE,

Burroughs CUBE, CDC Coop, Digital DECUS, and the Mark IV software package.[2]  I started counting women's names and men's names, first dozens, then hundreds, and soon enough thousands.  The accompanying committee reports and meeting programs often identified people as Mr. or Junior, Mrs., or Miss.  It was reasonably easy, though time-consuming, to gender identify 75% or more of these early computing specialists.  Here was new data to fill in the missing "white space" of women in early computing, from the late 1940s through the 1960s.

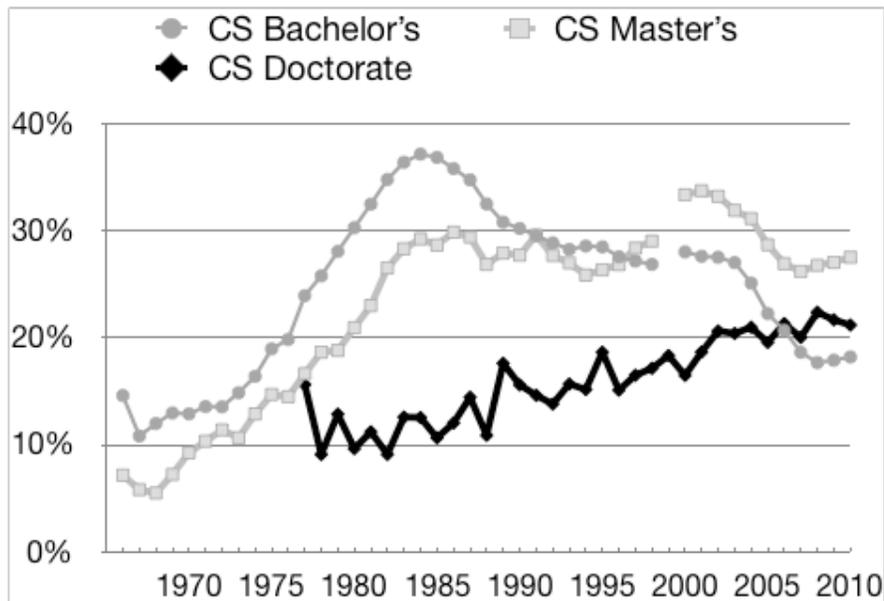

Figure 2: Women's share of Computer Science degrees (1966-2010)

Source: National Science Foundation and National Center for Science and Engineering Statistics, *Science and Engineering Degrees: 1966–2010: Detailed Statistical Tables* NSF 13-327 (Arlington: NSF, 2013), table 33, at www.nsf.gov/statistics/nsf13327/.

The underlying significant problem that this research addresses is the chronic underrepresentation of women in computing.  National statistics in the US (NSF and BLS) clearly indicated *growing* women's participation in computing from the mid-1960s to a peak in the mid-1980s, when women collected 37% of US CS bachelor's degrees (see figure 2) and then also constituted 38% of the white-collar IT workforce. In 1997, Tracy Camp published "The Incredible Shrinking Pipeline" in *Communications of the ACM*, where she noted that the problem of too-few women in computing was bad and getting worse.[3] Ever since, while the numbers have modestly improved after a longitudinal low point around 2008-9, clearly there are insufficient women in computing. The recent CRA–Taulbee report found that women collected 22.9% of North American undergraduate computer science degrees; it was back in *1977* (before the mid-1980s peak) when the NSF found this same figure.

---

[2] See CBI collections at https://cse.umn.edu/cbi/browse-cbia-collection-guides ; and Hagley Museum and Library's USE records at https://findingaids.hagley.org/repositories/3/resources/915

I continued my research hoping to find clues about what computing did "wrong" after the mid-1980s peak in women's participation and also to uncover what the nascent field clearly did "right" in attracting women into the computing field, greatly exceeding other technical fields such as engineering, medicine, and most of the physical sciences, which remained heavily male dominated for years. Evidence of women in computing abounds: Women attendees at SHARE meetings grew across 1955-73 to reach roughly 12%; women as SHARE's managers-officers went even higher, topping 25% by the late 1980s. In 1974 SHARE elected Shirley F. Prutch from Martin-Marietta Data Systems as its first female president; she became divisional vice president for systems integration in the 1980s and chaired a National Bureau of Standards panel on computer sciences and technology. Across the board, the user-group data indicated expanding women's participation in computing from the 1950s through the 1980s (see figure 3).

---

[3] Camp, T. "The Incredible Shrinking Pipeline," *Communications of the ACM* 40 no. 10 (October 1997): 103-110.

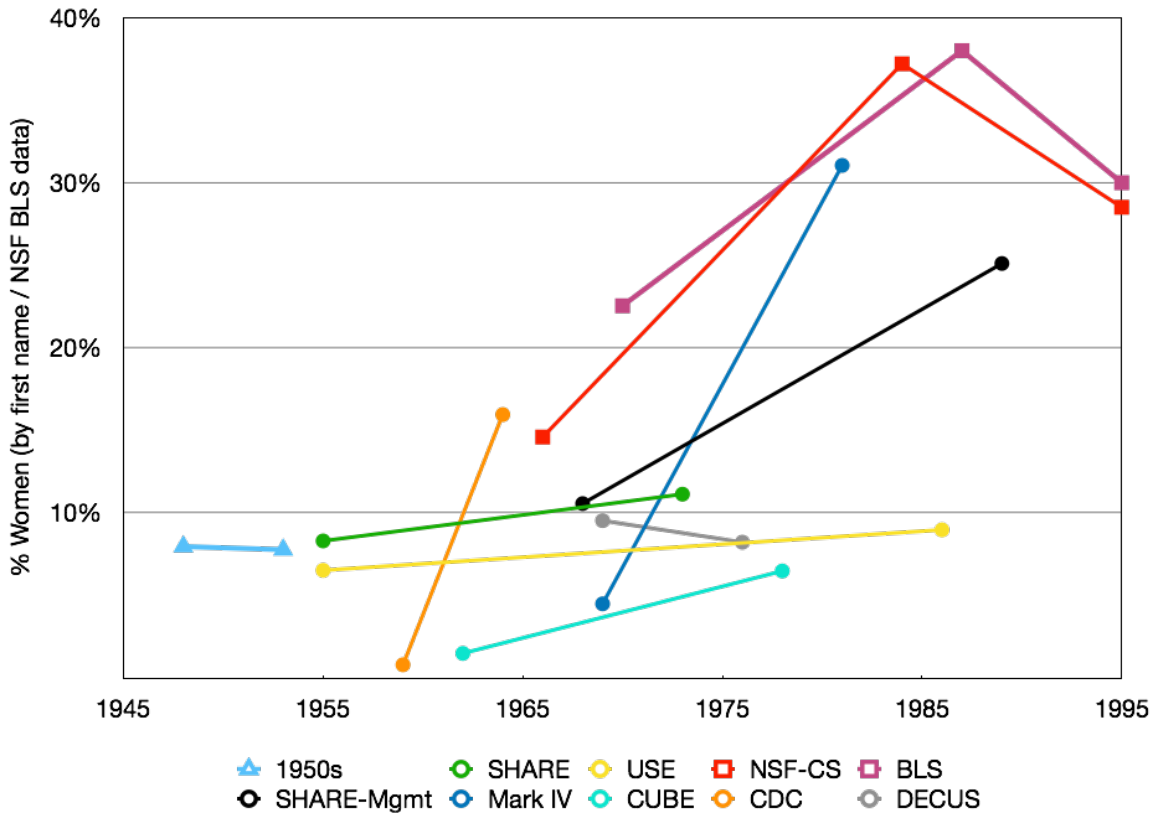

Source: Misa, "Dynamics of Gender Bias in Computing," *Communications of the ACM* 64 no. 6 (June 2021): 76-83

Prior to this research, the near-complete absence of quantitative data on women in computing before the 1970s had permitted some creative speculation about "when" and "why" women left computing. A conjecture about "making programming masculine" became established in the history of computing community. Scholars Jen Light and Nathan Ensmenger noted that in the 1950s there were many prominent women in computer programming (such as the ENIAC women, Grace Hopper, Ida Rhodes, Mina Rees, Florence Koons and many more), and the idea was hatched that women were at least 30% and possibly 50% of early computer programmers; these historians' figures were widely reported in newspapers, magazines, and Robin Hauser's acclaimed documentary "Code: Debugging the Gender Gap" (2015).[4] Women's seeming prominence didn't last;

---

[4] "By the 1960s, women made up 30% to 50% of all programmers, according to [historian Nathan] Ensmenger" (specifically citing the Robin Hauser film) according to Jane Porter, "The Fascinating Evolution of Brogramming And The Fight To Get Women Back," *Fast Company* (20

before long, according to "making programming masculine," with the professionalization of computing in the 1960s–1980s, male computer scientists, seeking to build their professional status, re-created computing as a male-centered (and, not coincidently, female-unfriendly) field.[5]

The "making programming masculine" conjecture [or MPM], posits that women left computing under the pressure of male-dominated professionalization in the 1960s and 1970s. The conjecture is based on a sociological theory of professionalization, but it is not supported by systematic or longitudinal data about the computing field during these years. Indeed, there are serious evidential shortcomings in MPM: my user-group and membership data noted above locates women in the computing field at roughly 10% (during the 1950s) and then growing in the 1960s through the 1980s. Similarly the BLS and NSF data indicate growth in women's participation through the all-time peak in the mid-1980s. My recent research into large datasets formed from the DBLP dataset (see figure 4 below) and ACM SIG conferences also demonstrates growth in women's participation in computing beginning in the 1960s and 1970s (lending no support to the conjecture that women, supposedly, were being chased out of the computing field during these years). Possibly the MPM storyline was not shared with Jean Sammet, developer of FORMAC, participant in COBOL, founder of SIGSAM, chair of SIGPLAN, and ACM's first female president 1974-76.

---

October 2014); see p. 77 of Misa, "Dynamics of Gender Bias in Computing" (full reference below). See additional citations claiming 30–50% women and/or women leaving computing in the 1960s or 1970s in note 33 of Misa, "Gender Bias in Computing."

[5] "Over the course of the 1960s and 1970s [note these years], male computer experts were able to successfully transform the 'routine and mechanical' (and therefore feminized) activity of computer programming into a highly valued, well-paying, and professionally respectable discipline," Ensmenger wrote. "They did so by constructing for themselves a distinctively masculine identity in which individual artistic genius, personal eccentricity, antiauthoritarian behavior, and a characteristic 'dislike of activities involving human interaction' were mobilized as sources of personal and professional authority." See Nathan Ensmenger, "'Beards, Sandals, and Other Signs of Rugged Individualism': Masculine Culture within the Computing Professions," *Osiris* 30 (2015): 38-65 at doi.org/10.1086/682955 quote p. 38.

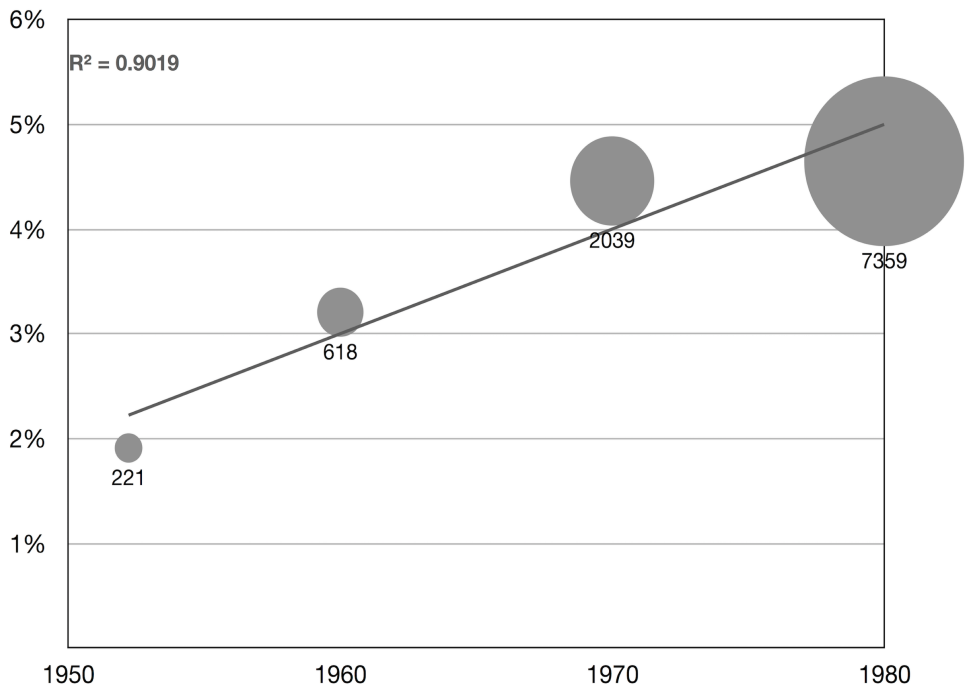

Figure 4: Women's participation in DBLP dataset (1950–1980)

My research method has significantly evolved since my early counting of women's and men's names.  My two early papers tabulate women's participation in the computing community, based on identified women's and identified men's names (and accompanying committee reports); I simply set aside, as gender unidentified, any name, including 'initial-only' names, that did not have reasonably clear gender associations.[6]  It was digital humanities scholars Blevins and Mullen who alerted me that name–gender associations are not fixed but oftentimes change across time.  Their formulation of the "Leslie problem" points out that Leslie was around 1900 preponderantly a man's name, assigned 92% to male babies; that Leslie evolved by around 1950 to be equally used for male and female babies; and that since 2000 Leslie is firmly a female name (96% or greater).  They identified additional names, such as Addison, Jan, Kendall, Madison, Morgan, and

---

[6] So, for example, in Figure 1 above: I'd tabulated names like Erik, Samuel, and Donald as men; initial-only names, unless labeled "Junior" (such as WP Barber, Jr.), as gender-unidentified; and names such as Jean and Gertrude as women.

Sidney, that experienced significant change in their gender association since 1930.[7] Another illustration of the temporal changes in gender association of "Leslie" comes from my analysis of 475 articles published in the ACM DL (see figure 5): ACM authors named Leslie were preponderantly male in 1970, became roughly half male and half female around 2005, and by 2020 were predominantly female, as was Leslie in the underlying SSA dataset.

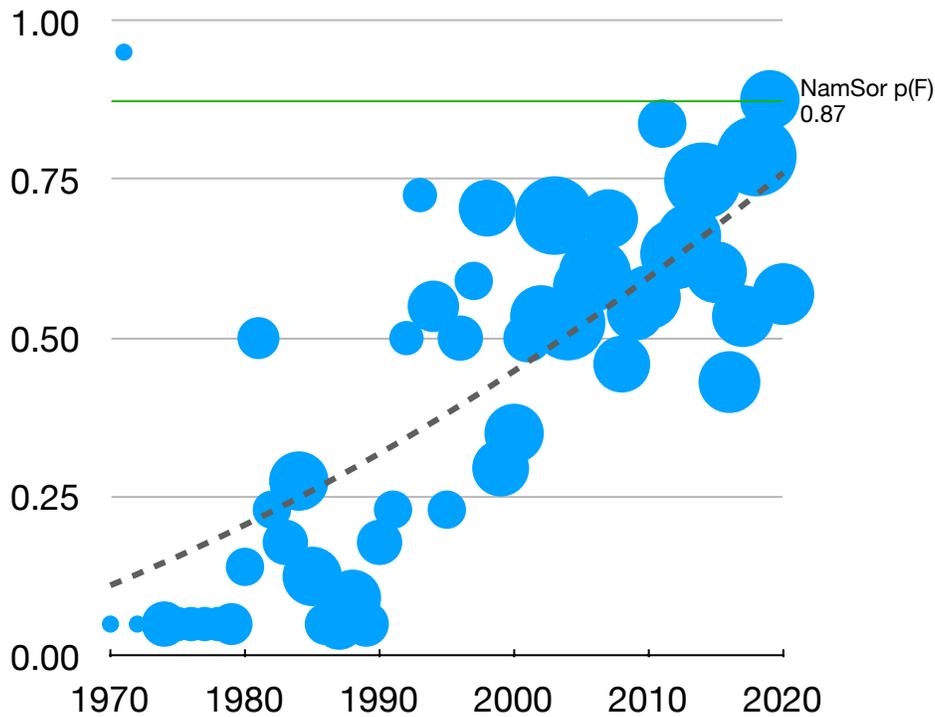

Figure 5: "Leslie problem" in ACM authors (1970–2020)

ACM authors named "Leslie" published 475 articles during 1970–2020. In 1970, there was one female Leslie (plotted as 0.95) and one male Leslie (0.05); in 1972-79, authors were all male (0.05); since then, the vertical plot is a weighted average of male and female Leslie's; gender not identified = 0.5. Note that gender ID software NamSor "predicts" p(F) for Leslie as 0.87, accurate for 2020 but inaccurate in earlier years.

My "Temporal Analysis and Gender Bias in Computing" available on ArXiv presents a systematic analysis of millions of first names (sample years: 1900, 1925, 1950, 1975, 2000) in the immense Social Security Administration dataset of first names assigned to male and female babies, year by year. While (say) George and Mary each have rather stable gender associations, strongly male and female, respectively; there are many names that are not stable: the name Johnnie (according to the SSA in 1960) was used to name 405 female babies and 1131 male babies, and so one can compute the "probability of a first name" (in a given year) [here: Johnnie in 1960 as p(F) = 0.26]. With this measure — different for each name, and changing across years — I identified 300 gender-unstable names, identifying the 25 names with the greatest gender shifts across 1925–1975 (see figure 6). Some names became more firmly associated with males, but overall I can identify and analytically measure a "net female shift."

---

[7] Cameron Blevins and Lincoln Mullen, "Jane, John . . . Leslie? A Historical Method for Algorithmic Gender Prediction," *Digital Humanities Quarterly* 9, no. 3 (2015): not paginated, at www.digitalhumanities.org/dhq/vol/9/3/000223/000223.html

Figure 6: Gender p(F) for top 24 gender-shift names [weighed by N] 1900-2000

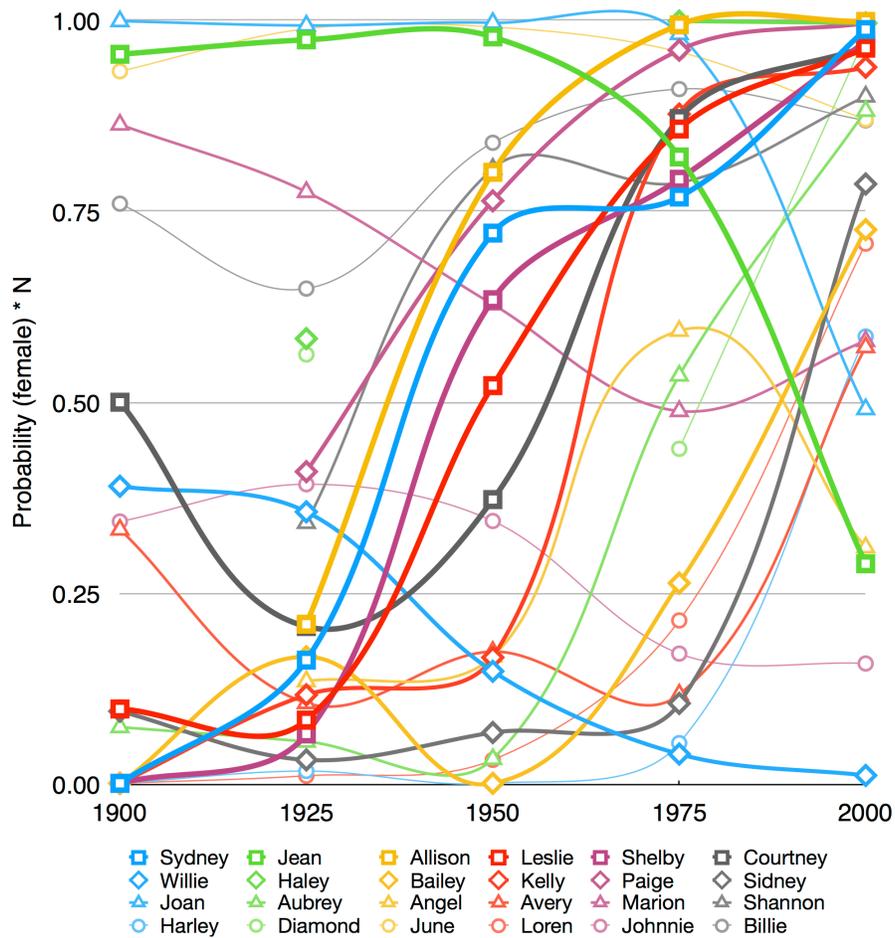

Source: Misa "Temporal Analysis and Gender Bias in Computing." *arXiv* (October 2022) <arxiv.org/abs/2210.08983>

With some datasets, I have employed extensive (and time consuming) lookups of individual computer scientists, using standard qualitative sources such as biographical dictionaries, the ACM and IEEE DLs, author websites, alumni lists, prize citations, and online lookups including obituaries. In the same way, I individually analyzed the 133 computer science Leslie's appearing in the ACM Digital Library 1970–2020, including Oxford's Leslie Ann Goldberg, MIT's Leslie Pack Kaelbling, and the Turing awardees Leslie Valiant and Leslie Lamport. Overall, like the general US population of Leslie's, the computer science Leslie's became more female over time approaching 2020 (see figure 5 above).

There's a further methodological wrinkle, in that very large *source populations* can be statistically "sampled" (using standard techniques) to result in *sample populations* that are still large but yet manageable for doing time-intensive qualitative analysis, yielding acceptable accuracies (say ±5%) and confidence levels (say 95%). For a source population of 600, a valid sample size (with this accuracy and confidence level) is calculated to be 238.[8] For the 1980 *source* population in DBPL (n=7,358 articles), I computed a slightly oversize *sample* to be 480 articles and then looked up all authors and co-authors for them. Finally, removing duplicate author-names resulted in 660 individual names to look up for 1980. That's a lot of qualitative spadework, and hours of internet searching; likely, different research methods, relying more heavily on scalable quantitative methods, will be needed for any much-larger computer-science populations.

My evolving research method is a classic "mixed method" with significant computer analysis of first names (based on targeted SSA lookups, attentive to year of publication) combined with qualitative analysis of anomalies, edge cases, and cultural factors. The "net female shift" in US names is significant since present-day name–gender associations are used by many gender identification software packages, including Gender-API, NamSor, and Genderizer.io. These work acceptably for first names today, but contribute to substantial error when used for first names in historical research. Reliance on Gender-API to identify thousands of research authors in the 1950s onward may contribute to the

---

[8] See Sharon L. Lohr, *Sampling: Design and Analysis* (Pacific Grove: Duxbury Press, 1999).

Women's Participation in Computing: Evolving Research Methods (version July 2024)anomalous findings (women are claimed to be 20% of CS authors in the 1950s, about ten-times higher than other published sources) recently reported by Lucy Wang and colleagues in *Communications of the ACM*.[9] Instead of these "present biased" software packages, I favor individual-name lookups in the immense SSA dataset, adjusted for year of publication (I have demonstrated analytically that a 'year shift' for SSA lookups of 30 years prior to year of publication results in the most accurate results).

Over the past six years, my research methods have evolved in several steps. Going forward, I can suggest using distinct research methods for different sized populations. For groups less than 100, there's much to recommend straight qualitative research using detailed personal investigations of individual members or researchers. Beyond about 100 and stretching up to about 500, I think a "mixed" method can be used, combining qualitative investigations of individuals supplemented with careful SSA lookups of name–gender associations (with appropriate year of gender-shift). Beyond this, and certainly past 1000 or so, it simply becomes exhaustively time-intensive to do qualitative personal research on the entire source population, and so judicious use of statistical sampling (combined with the above means for name identification) seems justified. The careful use of appropriately computed p(F) values permits the creation of inclusive datasets, including all the John's and Mary's as well as all the Leslie's, Johnnie's, Arie's and other names not rigidly used as male or female. Some persons have expressed wariness about any use of the SSA's gender-binary [F/M] data, which does not recognize today's sensibilities about gender practices; still, I believe we simply must use *some data* to investigate women's changing participation in computing.

This line of research has developed a reasonably robust research method that can be used to analyze both smaller and larger datasets. There are several substantive findings, too. I've measured gender bias used in "big data" analysis such as the gender-prediction software noted above (see first article in *Information and Culture* below). I can show quantitatively the uneven but steady grown in women's participation in early computing

---

[9] Lucy Lu Wang, Gabriel Stanovsky, Luca Weihs, and Oren Etzioni, "Gender Trends in Computer Science Authorship," *Communications of the ACM* 64 no. 3 (March 2021): see p. 83 figure 4 at doi.org/10.1145/3430803.

(computer user groups) and computer science (using the DBLP and ACM datasets) in a second article published in *Information and Culture*. I've indicated the likely consequences of a "net female shift" in US names for the gender study of computing over decades (available in ArXiv). The research I presented at the 2022 ACM SIG history seminar demonstrated striking differences within the ACM SIG communities, with SIG's varying from less than 10% women research authors to upwards of 40%. I've suggested that reform efforts seeking to improve women's participation should, in addition to the common focus on "computer science" as a unitary field, instead recognize profound differences within computer science as shown in SIG culture and organization (the ACM History Committee efforts to collect SIG organizational materials is much appreciated). ACM SIGs, of course, were often in cooperation or competition with independent technical conferences (such as AAAI spun off from SIGAI, or ISSAC from SIGSAM) as well as by conferences sponsored by the IEEE and IEEE Computer Society.[10] Using the IEEE DL, I am currently collecting data on women's participation in these additional regions of computer science and computer engineering.

---

[10] While collecting data on the "Annual Symposium on Switching and Automata Theory," or SWAT, suddenly, in the early 1970s, there appeared Juris Hartmanis, Robert E. Tarjan, and other luminaries in CS theory; in 1975 SWAT became the "Annual Symposium on Foundations of Computer Science," soon connected to IEEE and likely competing with ACM's SIGACT, which had 152 conference papers in 1990 but only 107 in 2000. Year by year data collected on the IEEE Computer Society's International Conference on Software Engineering (ICSE) across 1976 to 2010 reveal several distinct phases in women's participation as research authors. An article on ICSE is in preparation.